\documentclass[a4paper,12pt]{article}
\usepackage{amssymb,amsmath,amsthm,mathrsfs}
\usepackage{caption}
\usepackage{cite}
\usepackage{hyperref}
\usepackage{authblk}
\usepackage{tikz,pgfplots}
\usetikzlibrary{arrows,positioning,calc,fadings,decorations.pathreplacing,decorations}
\def\R{\mathbf{R}}
\def\B{\mathcal{B}}
\def\K{\mathcal{K}}
\newcommand\G[1]{\Gamma(#1)}
\newcommand\ads{\mathcal{M}}
\newcommand\eq[1]{(\ref{#1})}
\newcommand\C{\mathscr{C}^+}
\newcommand\rt{\longrightarrow}
\def\M{\mathcal{M}}
\title{Bulk reconstruction in AdS and Gel'fand-Graev-Radon Transform}
\author{Samrat Bhowmick\thanks{email: tpsb5@iacs.res.in}~} 
\author{Koushik Ray\thanks{email: koushik@iacs.res.in}}
\affil{\normalsize Department of Theoretical Physics, \authorcr Indian Association for the Cultivation of Science,\authorcr Kolkata 700 032. India.}
\author{Siddhartha Sen\thanks{email: siddhartha.sen@tcd.ie, sen1941@gmail.com}}
\affil{{\small CRANN}, \normalsize Trinity College Dublin, Dublin -- 2, Ireland}
\begin{document}
\maketitle
\begin{abstract}
\noindent
The bulk reconstruction formula for a Euclidean anti-de Sitter space 
is directly related to the inverse of the 
Gel'fand-Graev-Radon transform. Correlation functions of a conformal
scalar field theory in the boundary are thereby related to 
correlation functions of a self-interacting scalar 
field theory in the bulk at different loop orders. 
\end{abstract}
\setcounter{page}{0}
\thispagestyle{empty}
\clearpage
\noindent 
Holographic duality is a correspondence relating a quantum field theory
on a given space-time, referred to as the bulk,  
to one on its boundary, possibly 
in different ranges of couplings. The conjectured
duality between a theory of closed strings in the 
$n$-dimensional anti-de Sitter space  as the bulk and a 
conformal field theory 
on the boundary is an example of holographic
duality \cite{Maldacena:1997re,Witten:1998qj,Gubser:1998bc}.
It relates observables in the bulk anti-de Sitter space at weak coupling 
to correlation functions of a strongly coupled theory on the boundary and
vice versa.
The symmetries of a dual pair thus related ensure that both theories  
have the same number of degrees of freedom, although the scheme of
organizing them into fields  are rather 
different in the two theories. While the duality is
expected to work both ways as an equivalence relation, in practice, 
obtaining the theory in a given bulk 
from a given theory on the boundary appears to be more difficult.
This is known as the problem of bulk-reconstruction. 

Within the scope of scalar field theories both in the bulk and on the
boundary, to which we restrict ourselves in this note, a bulk-reconstruction 
procedure has been invented
\cite{Balasubramanian:1999ri,Bena:1999jv,Hamilton:2005ju,Hamilton:2006az,Hamilton:2006fh,
Kabat:2012hp,Kabat:2013wga,Kabat:2015swa,Roy:2015pga,Kabat:2017mun,Sanches:2017xhn,Goto:2017olq,Roy:2017hcp,Sarosi:2017rsq
}. 
While the normalizable modes of the bulk scalar fields are identified with
operators of the conformal field theory on the boundary, 
the latter produce bulk fields via an integral kernel, called the smearing
function.

In this note we relate the bulk reconstruction problem to integral geometry.
A scalar field on the boundary is taken to be the Gel'fand-Graev-Radon 
transform of a scalar field in the anti-de Sitter space
\cite{gel1966generalized}. This is a generalization of the Radon
transform on Euclidean spaces to Lobachevskian spaces.
Integral geometry studies the problem of
determination of a function on a manifold from the integral of the function 
on a family of submanifolds. For the
anti-de Sitter space, realized through a quadratic form in the
Euclidean space of one higher dimension,
a suitable choice of the submanifolds is obtained via
the null cone. The boundary is obtained as a limiting submanifold of the 
family. 
We show that if the Gel'fand-Graev-Radon transform of a function 
possesses certain scaling properties on the null cone, then it can be used to
write the function in the anti-de Sitter space from its integral on the 
boundary. Distributions or fields can be treated similarly. 

The invertible Gel'fand-Graev-Radon transform induces 
a transform between scalar field actions in the bulk and on the boundary,
namely, the two actions, while expressed in terms of different fields,
are numerically equal. 
We obtain the induced action on the boundary corresponding to a
self-interacting scalar field theory in the $n$-dimensional
bulk, in particular, a $\phi^k$-theory. The construction guarantees
that the theory on the boundary is conformal. Writing the corresponding
generating functionals then leads to relating correlation functions of the
two theories. 

Let us start by recalling some aspects of the Gel'fand-Graev-Radon (GGR)
transform. 
The most studied arena for Radon transform is manifolds of
constant curvature. 
Distributions on Grassmannian submanifolds of 
different codimensions are obtained through the Radon transform, which can
then be inverted \cite{helgason}. One example of such submanifolds is 
the set of geodesics \cite{solanes}. 
We shall restrict to another one-dimensional
submanifold, the set of lines through the origin. We begin with a discussion
of some features of it to be used here. 

Let $\ads$ denote the $n$-dimensional Lobachevskian space or the Euclidean 
anti-de Sitter space. In this note $n>1$.
In the coordinates $\{X^a| a=0,\cdots,n\}$ of the 
$(n+1)$-dimensional affine space $\R^{n+1}$
with metric $\eta_{ab}=\text{diag}(-1,1,\cdots, 1)$, $\ads$ is defined as 
the hypersurface 
\begin{equation}
\label{eq:ads}
\eta_{ab}X^aX^b = -L^2,
\end{equation}
where $L$ is a real number.
An equivalent description of $\ads$ is  
as the set of straight lines passing through the
origin of $\R^{n+1}$ within the region 
\begin{equation}
\eta_{ab}X^aX^b < 0.
\end{equation}
The isomorphism between these two stems from the fact that each of these lines
intersects the hypersurface \eq{eq:ads} at a single point. 
This is depicted in Figure~\ref{fig:ads2} for $n=2$.
We shall also consider the 
$n$-dimensional positive null cone $\C_n$ defined as the set of null
vectors $\xi^a$ in $\R^{n+1}$, that is,
\begin{equation}
\label{cone}
\eta_{ab}\xi^a\xi^b=0,\quad\xi^0\geqslant 0,
\end{equation}
also shown in Figure~\ref{fig:ads2}.
For a point $X$ in $\ads$ and a point $\xi$ in $\C_n$, let
us consider the family of hypersurfaces $\mathscr{S}_{n-1}(p)$, given by
\begin{equation}
\label{eq:H}
\Sigma :=
\eta_{ab} X^a\xi^b +p=0,
\end{equation}
where $p$ is a real parameter and $X^a$ and $\xi^a$ satisfy \eq{eq:ads} and
\eq{cone}, respectively. 
For a fixed non-zero $p$ this is called a
horosphere \cite{gel1966generalized}. The only solution for $X$ when $p$
vanishes is $X^a=\xi^a$, points on the cone. 
Looked upon as a subspace of the cone, this is
depicted in Figure~\ref{fig:ads}. Every line on the cone passing through the
apex intersects $\mathscr{S}_{n-1}$ only once. The hypersurface
$\mathscr{S}_{n-1}$ can alternatively viewed as a subspace of $\M$. Using the
isomorphism of $\M$ with the lines through the origin, these two descriptions
coincide as $X^0\longrightarrow\infty$. Hence the boundary of $\M$ falls on
the cone $\C_n$.
\begin{figure}[h]
\centering
\begin{minipage}{.48\textwidth}
\begin{tikzpicture}
\begin{axis}[axis lines = center, ticks=none,
 domain=0:5,
    y domain=0:2*pi,
    xmin=-5,
    xmax=5,
    ymin=-5,
    ymax=5,
    zmin=0,
    zmax=10,xlabel={$\scriptstyle X^1$},ylabel={$\scriptstyle X^2$},
zlabel={$\scriptstyle X^0$},
every axis x label/.style={at={(rel axis cs:0,0.5,0)},anchor=south},
        every axis y label/.style={at={(rel axis cs:0.5,0,0)},anchor=north},
        every axis z label/.style={at={(rel axis
cs:0.4,0.5,1.0)},anchor=west},
samples=25
]
\addplot3[surf,shader=interp,domain=0:6,y 
domain=0:1.21*pi]({.7*x*cos(deg(y))},{.7*x*sin(deg(y))},{sqrt(x^2+1)});
\addplot3[surf,shader=interp,opacity=.4,domain=0:6,y
domain=0:1.3*pi]({x*cos(deg(y))},{x*sin(deg(y))},{x});
\addplot3[domain=0:6]({.3*x*cos(deg(215))},{.5*x*sin(deg(215))},{x});
\addplot3[domain=0:6]({.2*x*cos(deg(65))},{.1*x*sin(deg(155))},{x});
\addplot3[domain=0:6]({.5*x*cos(deg(175))},{.6*x*sin(deg(235))},{x});
\addplot3[domain=0:6]({.7*x*cos(deg(215))},{.2*x*sin(deg(195))},{x});
\addplot3[domain=0:6]({.7*x*cos(deg(165))},{.5*x*sin(deg(185))},{x});
\addplot3[surf,shader=interp,domain=0:5,y
domain=1.21*pi:2*pi]({.7*x*cos(deg(y))},{.7*x*sin(deg(y))},{sqrt(x^2+1)});
\addplot3[surf,shader=interp,opacity=.4,domain=0:5,y
domain=1.3*pi:2*pi]({x*cos(deg(y))},{x*sin(deg(y))},{x});
\node at (rel axis cs:0.96,0.64,0.70) {$\scriptstyle\C_n$};
\node at (rel axis cs:0.37,0.28,0.62) {$\scriptstyle\text{AdS}_n$};
\end{axis}
\end{tikzpicture}
\caption{Null cone and $\text{AdS}_n$}
\label{fig:ads2}
\end{minipage}
\begin{minipage}{.48\textwidth}
\begin{tikzpicture}
\begin{axis}[axis lines = center, ticks=none,
 domain=0:5,
    y domain=0:2*pi,
    xmin=-5,
    xmax=5,
    ymin=-5,
    ymax=5,
    zmin=0,
    zmax=10,xlabel={$\scriptstyle X^1$},ylabel={$\scriptstyle X^2$},
zlabel={$\scriptstyle X^0$},
every axis x label/.style={at={(rel axis cs:0,0.5,0)},anchor=south},
        every axis y label/.style={at={(rel axis cs:0.5,0,0)},anchor=north},
        every axis z label/.style={at={(rel axis
cs:0.4,0.5,1.0)},anchor=west},
samples=25
]
\addplot3[surf,shader=interp,opacity=.4,domain=0:6,y
domain=0:1.3*pi]({x*cos(deg(y))},{x*sin(deg(y))},{x});
\addplot3[green!50!white,samples=60,line width=1pt,domain=0:2*pi](
{3*3*cos(deg(x))/(3+cos(deg(x))+sin(deg(x)))},
{3*3*sin(deg(x))/(3+cos(deg(x))+sin(deg(x)))},
{3*3/(3+cos(deg(x))+sin(deg(x)))});
\foreach \t in {1,...,9}{%
\addplot3[domain=0:6]({x*cos(deg(40*\t))},{x*sin(deg(40*\t))},{x});
}%
\addplot3[surf,shader=interp,opacity=.4,domain=0:6,y
domain=1.3*pi:2*pi]({x*cos(deg(y))},{x*sin(deg(y))},{x});
\node at (rel axis cs:0.95,0.5,0.70) {$\scriptstyle\C_n$};
\node at (rel axis cs:0.37,0.28,0.65) {$\scriptstyle\mathscr{S}_{n-1}$};
%
\end{axis}
\end{tikzpicture}
\caption{Null cone and $\mathscr{S}_{n-1}$}
\label{fig:ads}
\end{minipage}
\end{figure}
For a point $\xi$ on the cone $\C_n$, the GGR transform of a smooth 
function $f$  with bounded support on $\ads$ is given by the integral
\begin{equation}
\label{radon}
h_p(\xi)=\int_{\text{$\ads$}} f(X) \delta\left(\eta_{ab}X^a\xi^b+p\right) d^nX,\quad
\end{equation}
where 
$d^nX = \frac{dX^1 dX^2 \cdots dX^n}{X^0}$ is the volume
element on $\ads$ induced by \eq{eq:ads}.
The inverse transform, when exists, yields a function at 
a point in $\ads$ from a function on the light cone as
\begin{equation}
\label{irad}
f(X) = c_n\int_{\C_n} 
\frac{h_p(\xi)}{\left|\eta_{ab}X^a\xi^b + p\right|^n} d^n\xi,
\end{equation}
where $c_n$ is a constant, depending on the dimension $n$ and
$d^n\xi$ denotes the volume element on the null cone $\C_n$ induced by
\eq{cone}, namely,
\begin{equation}
\label{xivol}
d^n\xi = \frac{1}{\xi^n} d\xi^0 \cdots d\xi^{n-1}.
\end{equation} 
Consistency of \eq{radon} and \eq{irad} requires
\begin{equation}
\label{ads-delta}
c_n\int_{\C_n} \frac{\delta\left(\eta_{ab}X^a\xi^b+p\right)}{%
\left|\eta_{ab}Y^a\xi^b + p\right|^n} d^n\xi = \delta_{\text{$\ads$}}(X-Y),
\end{equation} 
where $\delta_{\text{$\ads$}}(X-Y)$ denotes the delta function 
on $\ads$. The constant $c_n$ is determined from the normalization of the 
delta function as
\begin{equation}
\label{cn}
c_n = {2L\alpha^2} \frac{\sin\frac{n\pi}{2}}{(2\sqrt{\pi})^{n-1}} 
\frac{\G{n}}{\G{\frac{n+1}{2}}}, 
\end{equation} 
where we have introduced a dimension-less constant $\alpha = p/L$. 
Computation of the constant is relegated to the end.

The GGR transform \eq{radon}
of a function and its inverse \eq{irad}
pertain to $\ads$ and the null cone. 
In order to relate it  to the bulk  reconstruction we need to first 
specify the boundary of $\ads$ and relate it
to the null cone. Let us consider an affine chart on $\ads$, 
\begin{gather}
\label{Xpara}
X^0 = \frac{zL}{2}\left(1+\frac{1+x^2}{z^2}\right), \quad
X^i = \frac{x^iL}{z}, \quad
X^n = \frac{zL}{2}\left(1-\frac{1-x^2}{z^2}\right), \\
x^2 = \sum_{i=1}^{n-1} (x^i)^2,\quad
\quad -\infty < x^i < \infty, \, 0\leqslant z < \infty;\quad
i=1,\cdots,n-1.
\end{gather}
These solve \eq{eq:ads}. The metric $g$ on $\ads$ is then given by
\begin{equation}
\label{metric}
ds^2 = \frac{L^2}{z^2}\left(dz^2+\sum_{i=1}^{n-1} (dx^i)^2 \right),
\end{equation}
the resulting volume element being
\begin{equation}
\label{dnX}
d^nX = \sqrt{g}dz d^{n-1}x = \frac{L^n}{z^n}dz d^{n-1}x.
\end{equation} 
In this chart the boundary $\B_{n-1}$ that we shall be 
concerned about is located at $z=0$, which in turn leads to $X^0\rt\infty$. 
The null cone $\C_n$ is a metric cone
$\R_+\times_{\xi^0}\mathbf{S}^{n-1}$ over an
$(n-1)$-dimensional sphere $\mathbf{S}^{n-1}$ with chart
\begin{gather}
\label{xipara}
\xi^i = \frac{2 \tilde x^i}{1+\tilde{x}^2} \xi^0, 
\quad\xi^n = - \frac{1-\tilde{x}^2}{1+\tilde{x}^2} \xi^0 \\
\tilde{x}^2= \sum_{i=1}^{n-1} (\tilde{x}^i)^2,\quad
-\infty < x^i < \infty,\, 0\leqslant\xi^0 < \infty;
\quad i=1,\cdots,n-1.
\end{gather}
These solve \eq{cone}.
In this chart the volume element \eq{xivol} takes the form 
\begin{equation}
\label{dxi}
d^n\xi = \frac{2^{n-1} (\xi^0)^{n-2}}{\left(1+\tilde{x}^2\right)^{n-1}} d\xi^0 \, d^{n-1} \tilde x \;.
\end{equation}
Substituting \eq{Xpara} and \eq{xipara} in \eq{eq:H} we obtain the equation
for $\mathscr{S}_{n-1}(p)$
in terms of the affine coordinates as
\begin{equation}
\label{bdry}
\left(z^2+\sum{(x^i-\tilde x^i)^2}\right)\xi^0L = 
zp\left(1+\sum{(\tilde x^i)^2}\right).
\end{equation} 
In the limit $\xi^0\rt\infty$ and $z\rt 0$, this leads to 
$x^i\rt\tilde{x}^i$, describing the boundary. 
Let us note that vanishing of $p$ results in
$c_n=0$, by \eq{cn}, consistent with the fact that there is no ``bulk" in this
limit.
In the following, we shall consider integration on the cone, as in \eq{irad}.
For such purposes, it is important to observe that the sphere at the base of the
cone admits smooth deformation to the hypersurface given by a constant 
value of $\xi^0$ and so does the boundary $\B_{n-1}$ too.

Let us assume that the previous considerations hold good for quantum fields. 
Let $\tilde\phi(\tilde{x})$ be a conformal field of dimension 
$\Delta$ on $\B_{n-1}$, whose coordinates are taken to be $\tilde{x}$. Then,
\begin{equation}
\label{conf:D}
\tilde\phi(\lambda\tilde{x}) = \lambda^{-\Delta}\tilde\phi(\tilde{x}),
\end{equation} 
where $\lambda=\lambda(\tilde{x})$ is any function on $\B_{n-1}$.
Let us assume that the function \eq{radon} is given by the conformal field as
\cite{Rychkov:2016iqz}
\begin{equation}
\label{cft:assum}
\begin{split}
h_p(\xi) &= h_p(\xi^0,\cdots,\xi^{n-1})\\
&\stackrel{def}{=}\tilde{\phi}\left(
\frac{2 \tilde x^1}{1+\tilde{x}^2} \xi^0, 
\frac{2 \tilde x^2}{1+\tilde{x}^2} \xi^0, \cdots,
\frac{2 \tilde x^{n-1}}{1+\tilde{x}^2} \xi^0
\right) \\
&=\left(\frac{2\xi^0}{1+\tilde{x}^2}\right)^{-\Delta} 
\tilde{\phi}(\tilde{x}),
\end{split}
\end{equation} 
where we have used \eq{xipara} at the second step.
Inserting this and \eq{dxi} in \eq{irad} we obtain 
\begin{equation}
 \label{ads2cone}
\phi(z,x) = \phi_0(n,\Delta)\int \mathcal{K}(z,x|\tilde x)
\tilde{\phi}(\tilde x) \, d^{n-1}\tilde x \;,
\end{equation}
with
\begin{equation}
 \label{smrfunc}
 \mathcal{K}(z,x|\tilde x) = 
{\left(\frac{z^2+\sum{(x^i-\tilde x^i)^2}}{z}\right)^{\Delta+1-n}}\;,
\end{equation} 
where we denoted the field in $\ads$ as $f(X)=\phi(z,x)$. The
constant $\phi_0(n,\Delta)$ is given by
\begin{equation}
\phi_0(n,\Delta) = \frac{2^{n-1-\Delta}c_n}{\alpha^{1+\Delta}L^n}
\int_0^{\infty} \frac{\zeta^{n-2-\Delta}}{|1-\zeta|^n}d\zeta,
\end{equation} 
where we have defined
\begin{equation}
\zeta = \left(\frac{\xi^0}{z\alpha}\right)\frac{z^2+(x-\tilde{x})^2}{1+\tilde{x}^2}.
\end{equation} 
The integral in $\zeta$ can be evaluated using the formula
\begin{equation}
\label{Barnes}
\int_0^{\infty} \frac{x^a}{(1-x)^n}dx = \frac{\G{a+1}\G{1-n}}{\G{a-n+2}},
\end{equation} 
which, in turn, can be obtained by writing the denominator of the integrand
as a Barnes' integral. 
This yields
\begin{equation}
\phi_0(n,\Delta) = \frac{2^{n-\Delta-1}c_n}{\alpha^{\Delta+1}L^n} 
\frac{\pi}{\sin n\pi}\frac{\G{\Delta+1}}{\G{n}\G{\Delta+2-n}}
(1+(-1)^n).
\end{equation}
Both the expression for $\phi_0$ and the formula \eq{Barnes} are singular,
as written, since $n$ is an integer and $n>1$. In order to obtain the
normalized GGR transform these are to be evaluated in a regularized manner.
Using the formula $\Gamma(z)\Gamma(1-z)=\pi/\sin\pi z$,  
plugging in the value of $c_n$ from \eq{cn} and further using the regularized
expression
\begin{equation}
\label{reg1}
(1+(-1)^n) \frac{\sin\frac{n\pi}{2}}{\sin n\pi} = e^{in\pi/2}
\end{equation} 
we finally obtain
$\phi_0(n,\Delta)=\Phi_0(n,\Delta)/L^{n-1}$, with
\begin{equation}
\label{phi0:odd}
\Phi_0(n,\Delta)=\frac{e^{in\pi/2}}{(2\alpha)^{\Delta-1}\pi^{(n-3)/2}}
\frac{\G{\Delta+1}}{\G{\Delta-n+2}\G{\frac{n+1}{2}}}.
\end{equation} 
Using this expression
in \eq{ads2cone} gives the formula for bulk reconstruction of a
conformal field of scaling dimension $\Delta$ from the boundary of an anti-de
Sitter space of dimension $n$ \cite{Hamilton:2005ju}.
The expression \eq{reg1} requires qualification. The inversion of the
Radon transform is a well-known ill-posed problem. It involves computing 
integrals with prescribed regularization to determine the constant $c_n$ 
\cite{gel1966generalized}. In the present case, the assumption of
conformality, \eq{cft:assum} brings in factors which conspire to cancel the
singularities, yielding the non-singular expression \eq{reg1}.

The bulk field $\phi$ expressed as the
inverse transform \eq{ads2cone} when operated on by the
Laplacian $\Box_{\M}$ on $\M$ obeys the equation
\begin{equation}
\label{eom1}
\Box_{\M}\phi(z,x)=\frac{1}{\sqrt g} \partial_\mu \left( 
\sqrt g g^{\mu\nu} \partial_\nu\phi \right) = \frac{\Delta(\Delta-n+1)}{L^2}\phi(z,x),
\end{equation}
implying that the scalar field $\phi$ is massive, with mass $m$ given by 
\begin{equation}
 \label{mass}
 m^2 = \Delta(\Delta-n+1)/L^2.
\end{equation}

Now that we have obtained the bulk reconstruction formula as a transform which
is invertible, we can use it to induce actions from the bulk to the boundary
and vice versa. For example, using the metric \eq{metric}, 
the action of a free scalar field in $\M$ is 
\begin{equation}
\label{action:bl}
S(\phi) = \int d^{n-1}x dz \sqrt{g} \left[g^{\mu\nu}\partial_{\mu}\phi(z,x)\partial_{\nu}\phi(z,x) +m^2 \phi^2(z,x) \right].
\end{equation} 
Plugging in \eq{ads2cone} with \eq{phi0:odd} in this action we obtain the
action on the boundary. 
From the first term of the (\ref{action:bl}), we obtain
\begin{equation}
\label{action:bd}
\tilde{S}(\tilde{\phi}) =S(\phi)
= \frac{\Phi_0^2(\Delta+1-n)^2 }{L^n}\int\mathcal{P}(\tilde{x},\tilde{x}')
\tilde{\phi}(\tilde{x}) \tilde{\phi}(\tilde{x}')
d^{n-1}\tilde{x}d^{n-1}\tilde{x}',
\end{equation} 
where $\mathcal{P}$ involves integrations over $z$ and $x^i$, $i=1,\cdots,n-1$.
Let us emphasize that the two actions $S$ and $\tilde{S}$ live on different
spaces and contain different fields, but are numerically equal. This is a
consequence of the fact that the holographic relation between the fields
$\phi$ and $\tilde{\phi}$ has been expressed as an invertible transform.
The integration over $z$ can be evaluated using, say, 
\verb|Mathematica|, to obtain it in the form of a sum of terms like
\begin{equation}
\mathcal{P}(\tilde{x},\tilde{x}') \sim \int |x-\tilde{x}|^r |x-\tilde{x}'|^s 
G\left(\frac{|x-\tilde{x}|^2}{|x-\tilde{x}'|^2}\right) d^{n-1}x,
\end{equation} 
with $r+s = 2\Delta-3n+3$ and some function $G$. Exact expressions are given
in the end. Hence, using \eq{conf:D},
under the scaling $\tilde{x}\rt\lambda\tilde{x}$ accompanied by a change of
variables $x\rt \lambda x$, $\mathcal{P}$ scales as 
$\mathcal{P}(\tilde{x},\tilde{x}')\sim|\tilde{x}-\tilde{x}'|^{2(1+\Delta-n)}$, 
and the action $\tilde{S}$ remains invariant. Consequently, 
the action $\tilde{S}$ on the boundary can be written as
\begin{equation}
\label{Sbd}
\tilde{S} = \frac{P_0}{L^n}\int\frac{\tilde{\phi}(\tilde{x}) \tilde{\phi}(\tilde{x}')}{%
|\tilde{x}-\tilde{x}'|^{2(n-\Delta-1)}}
d^{n-1}\tilde{x}d^{n-1}\tilde{x}',
\end{equation} 
where $P_0$ depends on the mass.

Let us now consider a self-interacting scalar field in the bulk and 
derive relations between the correlation functions of the bulk and the 
boundary theories. Adding a potential $V(\phi)$ to the bulk 
action \eq{action:bl} we consider 
\begin{equation}
S_{I}(\phi) = S(\phi) + S_{int}(\phi),
\end{equation} 
where $S_{int}(\phi)=\int dzd^{n-1}x V(\phi)$.
The generating functional of the interacting theory,
\begin{equation}
Z_I = \int D\phi\ e^{S_I(\phi)},
\end{equation} 
can be expressed in terms of that of the
non-interacting theory plus a source term. Introducing a source $J$ in the
bulk we write
\begin{equation}
S(\phi, J) = 
S(\phi) +\int\sqrt{g} J(z,x) \phi(z,x) dz d^{n-1}x.
\end{equation} 
Then 
\begin{equation}
\label{ZI}
Z_I = \left. e^{S_{int}
\left(\frac{1}{\sqrt{g}}
\frac{\delta}{\delta J}\right)}
Z[J]\right|_{J=0},
\end{equation} 
where 
\begin{equation}
\label{ZJ}
Z[J] = \int D\phi\ e^{S(\phi,J)}.
\end{equation} 
Correlation functions are computed as moments by differentiating $Z[J]$
with respect to the source.
Using \eq{action:bd} and the transform \eq{ads2cone} we rewrite the
source term in $S(\phi,J)$ in terms of $\tilde{\phi}$ to obtain
\begin{equation} 
\tilde{S}(\tilde{\phi},\tilde{J})=\tilde{S}(\tilde\phi) 
+ \int \tilde J(\tilde x) \tilde \phi(\tilde x) d^{n-1} \tilde x,
\end{equation} 
where we have defined
\begin{equation}
\label{JtildeJ}
\tilde J(\tilde x) = \frac{\Phi_0(n,\Delta)}{L^{n-1}}
\int  \sqrt{g}\ \mathcal{K}(z,x|\tilde x) J(z,x) dz\ d^{n-1}x.
\end{equation} 
Derivatives with respect to the sources are related by 
\begin{equation}
\label{J:der}
\frac{\delta}{\delta J'(z,x)} \stackrel{def}{=}
\frac{1}{\sqrt{g}}
\frac{\delta}{\delta J(z,x)} = \frac{\Phi_0(n,\Delta)}{L^{n-1}}\int
d^{n-1}\tilde{x}\  
\mathcal{K}(z,x|\tilde x) \left(\frac{\delta}{\delta \tilde J(\tilde x)}
\right).
\end{equation} 
Equality of the actions $S$ and $\tilde{S}$ 
then implies that the correlation
functions computed in the bulk and in the boundary theories, respectively
as normalized $J$-moments of $Z[J]$ and $\tilde{J}$-moments of $\tilde{Z}$, 
with the latter defined as
\begin{equation}
\tilde{Z}[\tilde{J}] = \int D\tilde\phi\ e^{\tilde{S}(\tilde\phi,\tilde{J})},
\end{equation} 
are equal.

Let us illustrate this with an example by considering  the interaction
$V(\phi)=\lambda\phi(z,x)^k$,
for a fixed positive integer $k$.
The $\ell$-th loop contribution to the two-point function is 
\begin{equation} 
\label{CorInt}
\langle\phi(z_1,x_1)\phi(z_2,x_2)\rangle_{\ell}
= \frac{\lambda^\ell}{\ell\, !} 
\left.
\frac{\delta^2}{\delta J'(z_1,x_1)\delta J'(z_2,x_2)} 
\prod_{i=1}^{\ell}\int 
dz'_id^{n-1}y_i\frac{\delta^k}
{\delta J'(z'_{i},y_{i})^k}
Z[J]\right|_{J=0}.
\end{equation} 
Using $\tilde{Z}[\tilde{J}]$ in place of $Z[J]$ and \eq{J:der}, we obtain
\begin{equation} 
\begin{split}
\label{BBCor}
\langle \phi(z_1,x_1) \phi(z_2,x_2)\rangle_{\ell} = 
\frac{\lambda^{\ell} \Phi_0^{k\ell+2}}{L^{(k\ell+2)(n-1)}} 
\int d^{n-1}\tilde{x}_1 d^{n-1}\tilde{x}_2
\prod_{i=1}^{\ell} 
\prod_{j=1}^k
dz'_i d^{n-1}y_i d^{n-1}\tilde{y}_{(i,j)}\\
\K(z_1,x_1|\tilde{x}_1)\K(z_2,x_2|\tilde{x}_2)
\K(z'_i,y_i|\tilde{y}_{(i,j)})
\left\langle
\tilde{\phi}(\tilde{x}_1)\tilde{\phi}(\tilde{x}_2) 
\prod_{\imath=1}^{\ell} 
\prod_{\jmath=1}^k
\tilde{\phi}(\tilde{y}_{(\imath,\jmath)}) \right\rangle,
\end{split}
\end{equation} 
where 
\begin{equation}
\langle\tilde\phi(\tilde x_1)\tilde\phi(\tilde x_2)\cdots
\tilde\phi(\tilde x_{m}) \rangle 
=
\left.\prod_{i=1}^{m}
\frac{\delta}{\delta \tilde J(\tilde x_i)}
\tilde{Z}[\tilde{J}]\right|_{\tilde{J}=0}
\end{equation} 
denotes the $m$-point function of the boundary theory. 
This feature generalizes. Generally, the $\ell$-th loop contribution to the
$p$-point function in the bulk is determined by the $(k\ell+p)$-point function
of the boundary theory. 

To conclude, we have obtained the HKLL bulk reconstruction formula as the
Gel'fand-Graev-Radon transform on the $n$-dimensional Euclidean anti-de Sitter
space for $n>1$. This is achieved with the assumption \eq{cft:assum} that the
transform on the positive light cone is given by a conformal field on the
boundary.
This allows us to relate the scalar field actions in the
bulk and that at the boundary. While expressed in terms of the transformed
fields the actions are numerically equal. We use this to write the generating
functionals in the two theories, incorporating interactions in the bulk. It
is then showed that for a $\phi^k$ interaction term in the bulk
the $p$-point correlation function at the $\ell$-th loop is related to the
$(p+k\ell)$-point correlation function in the boundary, in keeping
with holography.

Integral geometry has been discussed earlier in the context of AdS-CFT
duality. In one approach the CFT was associated to a Fourier
transform \cite{tolfree}, which in turn was related to the Gel'fand-Graev
technique. In another approach  \cite{Czech:2016xec} OPE blocks on 
the kinematic space defined by 
the configuration space of two points on $\text{AdS}_3$,
was considered. The kinematic space, having double the dimension, is looked upon
as the space of geodesics in the bulk ending on the pair of points at the
boundary. Geodesic Radon transform was then used to interpret the OPE blocks
in the bulk as ``geodesic operators", relating the correlators in terms of
the length of geodesics in the bulk. The approach we have taken in here is 
a more direct one. We consider the boundary of $\text{AdS}_{n}$ and 
identify it on
the positive light cone. Then, an inverse Radon transform, in conjunction
with the assumption of conformality on the light cone \eq{cft:assum} is used
to obtain the bulk fields. The assumption of conformality on $\C_n$ restricts
the space of distributions on the boundary, which otherwise may be taken to
be the Schwartz space. Indeed, it is because of this assumption that the
otherwise ill-defined normalization constant $c_n$ is multiplied with an
extra term to make the coefficient of the transform finite, as in \eq{reg1}.
We believe that this computation will be useful in understanding the
structure of the bulk reconstruction procedure.

\subsection*{Computation of $c_n$ in \eq{cn}}
In order to compute the constant $c_n$ it is convenient to first write
\eq{ads-delta} after integrating on $X$ as
\begin{equation}
I=\int_{\M}\int_{\C_n} \frac{\delta\left(\eta_{ab}X^a\xi^b
+p\right)}{%
\left|\eta_{ab}Y^a\xi^b + p\right|^n} d^n\xi d^nX.
\end{equation} 
We shall not perform the integration over $X$, but keeping it in the integral is
useful to keep track of factors occurring in change of variables. 
Interpreting the integral as
\begin{equation}
I=\int_{\M}\left(
\int\limits_{\C_n\cap\mathscr{S}_{n-1}} \frac{1}{%
\left|\eta_{ab}Y^a\xi^b + p\right|^n} d^n\xi\right) d^nX.
\end{equation} 
we perform the integration over $\xi$ by restricting 
$d\xi^0 d^{n-1}\tilde{x}$ to
$\mathscr{S}_{n-1}$ as
$d^{n-1}\tilde{x}/\frac{\partial\Sigma}{\partial\xi^0}$, with $\xi^0$
evaluated in terms of $\tilde{x}$ from \eq{bdry}.
Moreover, we substitute the affine coordinates \eq{Xpara} for 
$X$ and $Y$ as $(z,x)$ and $(w,x')$, respectively along with \eq{dnX} in the
integral. We first define new variables as
\begin{equation}
z=w\tau,\quad x = w\hat{x}, \quad x'=w\widehat{x'}, \quad
\tilde{x}=w\widehat{\tilde{x}}.
\end{equation} 
Then with a further change of variables
\begin{equation}
\hat{x}=\widehat{\tilde{x}}+y,\quad
\hat{x}-\widehat{x'}=r,
\end{equation} 
we rewrite the integral as 
\begin{equation}
I = \frac{2^{n-1}}{L\alpha^2}\int\frac{(\tau^2+y^2)}{%
\left|\tau^2+y^2-\tau(1+(y-r)^2)\right|^n} 
\frac{d\tau}{\tau} d^{n-1}r d^{n-1}y.
\end{equation} 
Using the symmetries of $\M$ we now set $r=0$ in the integrand
\cite{gel1966generalized}, replacing
the integration over $r$ with the volume of the $(n-1)$-dimensional sphere,
$V_{n-1}=\frac{2\pi^{(n-1)/2}}{\G{(n-1)/2}}$.
Defining further, $y=\sqrt{\tau}\rho$, the integral finally assumes the form 
\begin{equation}
\begin{split}
I=\int_0^1\left[\frac{2^n\pi^{(n-1)/2}V_{n-1}}{\alpha^2L\G{(n-1)/2}}
\left(\int_0^1\frac{(1+\rho^2)\rho^{n-2}}{(1-\rho^2)^n}d\rho\right)
\left(\frac{(1+\tau)(1+\tau^{2n-2})}{%
(1-\tau)^n}\tau^{-(n+1)/2}\right)\right]d\tau,
\end{split}
\end{equation} 
Evaluating the integration over $\rho$ it can be seen that the integrand
inside the square braces is supported at $\tau=1$, which corresponds to
$z=w$. The integrand is thus a delta function, as in \eq{ads-delta}, with
strength $1/c_n$, with $c_n$ given by \eq{cn}. 
\subsection*{Constants appearing in the action}
The kernel  $\mathcal{P}(\tilde{x},\tilde{x}')$ in \eq{action:bd} is 
\begin{equation} 
\mathcal{P}(\tilde{x},\tilde{x}')
=\int
\left( P_1(x,\tilde{x},\tilde{x}')
+\sum_{i=1}^{n-1}(a^ib^i)P_2(x,\tilde{x},\tilde{x}')
+m^2P_3(x,\tilde{x},\tilde{x}') \right) d^{n-1}{x},
\end{equation} 
where we have defined $a^i=x^i-\tilde{x}^i$ and $b^i=x^i-\tilde{x}'^i$. The
three functions are written in terms of the Gaussian hypergeometric function
$F$ as 
\begin{equation}
\begin{split}
P_1(x,\tilde{x},\tilde{x}')
=& \scriptstyle a^{-2 n+2\Delta +2} b^{1-n}
\left({3+n^2-4n}\right)
\frac{\Gamma\left(\frac{n-3}{2}\right) 
\Gamma \left(\frac{n-2  \Delta -1}{2}\right)}{8\Gamma (n- \Delta)}
F\left(\frac{n-2  \Delta -1}{2},n- \Delta
   ;\frac{1-n}{2};\frac{b^2}{a^2}\right) 
\\
&+\scriptstyle \left(a^{-2n+2\Delta+2}b^{1-n}+a^{2\Delta-2n}b^{3-n}\right) 
\left({4n+2\Delta(n-3)-n^2-3}\right) 
\frac{\Gamma\left(\frac{n-3}{2}\right) 
\Gamma\left(\frac{n-2\Delta-1}{2}\right)}{8\Gamma(n-\Delta)}
F\left(\frac{n-2  \Delta +1}{2},n- \Delta
   ;\frac{3-n}{2};\frac{b^2}{a^2}\right) 
\\
&+\scriptstyle  a^{2\Delta-2n}b^{3-n} 
\left({n^2+4\Delta^2-4n\Delta-1}\right)
\frac{\Gamma\left(\frac{n-3}{2}\right)
\Gamma\left(\frac{n-2  \Delta -1}{2}\right)}{8\Gamma(n-\Delta)}
F\left(\frac{n-2\Delta +3}{2},n- \Delta;
\frac{5-n}{2};\frac{b^2}{a^2}\right)
\\
&-\scriptstyle \left(a^{-3n+2\Delta+1}b^{2}+a^{-3n+2\Delta+3}\right)
\left({3n^2-2\Delta-2n\Delta-3}\right)
\frac{\Gamma\left(-\frac{n+1}{2}\right) 
\Gamma\left(\frac{3(n-1)}{2}-\Delta\right)}{8\Gamma(n-\Delta)}
F\left(\frac{3n-2\Delta-1}{2},n- \Delta;
\frac{n+1}{2};\frac{b^2}{a^2}\right)
\\
& +\scriptstyle  a^{-3n+2\Delta+1}b^2 
\left({3+9n^2+4\Delta^2+8\Delta-12n\Delta-14n}\right)
\frac{\Gamma \left(-\frac{n+1}{2}\right) 
\Gamma\left(\frac{3 (n-1)}{2}-\Delta \right)}{8\Gamma(n-\Delta)}
F\left(\frac{3n-2\Delta +1}{2},n-\Delta;\frac{n+3}{2};\frac{b^2}{a^2}\right) 
\\
&-\scriptstyle a^{-3n+2\Delta+3}\left({n^2-1}\right)
\frac{\Gamma\left(-\frac{n+1}{2}\right)
\Gamma\left(\frac{3(n-1)}{2}-\Delta\right)}{8\Gamma(n-\Delta)}
F\left(\frac{3n-2\Delta-3}{2},n-\Delta;\frac{n-1}{2};\frac{b^2}{a^2}\right)
\end{split}
\end{equation}
\begin{equation} 
\begin{split}
P_2(x,\tilde{x},\tilde{x}')
 =&  a^{2  \Delta -3 n+1} 
\frac{\Gamma\left(\frac{1-n}{2}\right)
\Gamma\left(\frac{3n-2\Delta-1}{2}\right)}{2\Gamma(n-\Delta)} 
F\left(\frac{3 n-2  \Delta -1}{2},n- \Delta ;\frac{n+1}{2};
\frac{b^2}{a^2}\right)
\\
&+ a^{2\Delta -2 n}b^{1-n} 
\frac{\Gamma\left(\frac{n-2  \Delta +1}{2}\right) 
\Gamma\left(\frac{n-1}{2}\right)}{2 \Gamma (n- \Delta )} 
F\left(\frac{n-2\Delta +1}{2},n-\Delta;\frac{3-n}{2};\frac{b^2}{a^2}\right)
\end{split}
\end{equation} 
\begin{equation}
\begin{split}
P_3(x,\tilde{x},\tilde{x}')
 =& a^{2\Delta -3 n+3} 
\frac{\Gamma\left(\frac{3n-2\Delta-3}{2}\right)
\Gamma\left(\frac{1-n}{2}\right)}{2 \Gamma (n- \Delta -1)}
F\left(\frac{3n-2\Delta-3}{2} ,n- \Delta
   -1;\frac{n+1}{2};\frac{b^2}{a^2}\right)
\\
   &+a^{2  \Delta -2 n+2} b^{1-n}
\frac{\Gamma\left(\frac{n-2\Delta -1}{2}\right)
\Gamma\left(\frac{n-1}{2}\right)}{2\Gamma (n- \Delta -1)}
F\left(\frac{n-2\Delta -1}{2},n- \Delta -1;
\frac{3-n}{2};\frac{b^2}{a^2}\right)
\end{split}
\end{equation} 
The formulas are written in the least cluttered form. The Gamma functions are
to be analytically continued depending on the values of $n$ and $\Delta$.
\subsection*{Acknowledgements}
SB and KR thank Bobby Ezhuthachan for illuminating discussions 
on various occasions.

\end{document}